\documentclass[11pt,twoside]{article}

\usepackage{edwin}
\usepackage{epsf}
\usepackage{psfig}
\usepackage{lscape}
\usepackage{longtable}
\usepackage{latexsym}
\usepackage{amssymb,amsmath}
\usepackage{gplPts}

\def\GStar{\gplPt{Star}}
\begin{document}
\title{Use of Astronomical Literature - A Report on Usage Patterns}   
\author{Edwin A. Henneken\altaffilmark{1}, Michael J. Kurtz, Alberto Accomazzi, Carolyn S. Grant, Donna Thompson, Elizabeth Bohlen, Stephen S. Murray}
\affil{Harvard-Smithsonian Center for Astrophysics, 60 Garden Street, Cambridge, MA 02138}
\altaffiltext{1}{Corresponding author at: Harvard-Smithsonian Center for Astrophysics, 60 Garden Street, Cambridge, MA 02138, USA. {\it E-mail address}: ehenneken@cfa.harvard.edu (E. Henneken).}
\begin{abstract} 
In this paper we present a number of metrics for usage of the SAO/NASA Astrophysics Data System (ADS). Since the ADS is used by the entire astronomical community, these are indicative of how the astronomical literature is used. We will show how the use of the ADS has changed both quantitatively and qualitatively. We will also show that different types of users access the system in different ways. Finally, we show how use of the ADS has evolved over the years in various regions of the world.

The ADS is funded by NASA Grant NNG06GG68G.
\end{abstract}
\section{Introduction}

The SAO/NASA Astrophysics Data System (hereafter ADS), is a digital library and a vital source for bibliographic information in astronomy. The vast majority of astronomical researchers in the world use the ADS on a daily or near-daily basis. The use of the ADS has not only changed quantitatively but also qualitatively. Initially almost exclusively used by professional astronomers, the ADS now also has become a public service through external, general search engines (like Google, Yahoo, Microsoft Live Search and Ask.com, to name a few). In~\citet{henneken07} we observed that up to the
middle of 2004, the number of ADS users doubled on a bi-yearly basis. Since the ADS started to be indexed by general search engines, the number of incidental users has increased dramatically. However, the number of typical users (more than 10 visits per month) has continued to follow the same growth pattern. 

With different types of users come different types of use. A professional astronomer has different interests than an occasional user. One way of illustrating this is to look at the distribution of publication years for the literature people are interested in. We will also look at the diversity of ADS users from a geographical point of view. This will indicate whether increased Internet access actually results in an increase of ADS usage. This is particularly interesting with respect to aspects of the ``Digital Divide'' (see e.g.~\citet{ITU07}). In the next section, we will describe the character of the data we are working with. The following section will show the results, which will then be discussed in section 4. 

\section{Data}

The ADS is an electronic library where the system log files record queries and access to its records over time. For every bibliographic record in the ADS, a user can choose to view or access various types of metadata associated with that record. A ``visit'' (or ``read'') is defined as the selection of a metadata link. The ADS's data log entries record what type of data item was selected for which article. For example, in the period of January and February of 2008, 72\% of all requests were for an abstract, 19\% for the full article text, 4\% for citation histories and 3\% for e-prints from arXiv. Our data log entries show where a visitor came from (for example, whether a visitor used the ADS directly or came in via an external source) and what he/she read. When a user requests more information than just an abstract and/or performs a query, a cookie is assigned to that user. This allows us to compile frequency statistics for these users and determine the group of ``typical users''.

These ingredients provide us with the information we need to analyze the behavior of different types of users. For our analysis we use the ADS log files of January and February of 2008. These log files represent over 5.9 million requests from 317,753 unique users with a cookie and 1,071,416 unique users without a cookie, defined as the number of unique IP addresses associated to queries without a cookie. In the next section, besides looking purely at usage patterns,
we will also compare usage to Gross Domestic Product (GDP) and Internet usage. We used GDP data from the World Economic Outlook database (\citet{IMF08}). Internet usage was obtained from the EarthTrends database (\citet{WRI08}).

\section{Results}
\subsection{General readership}
\begin{figure}[!ht]
  \plotone{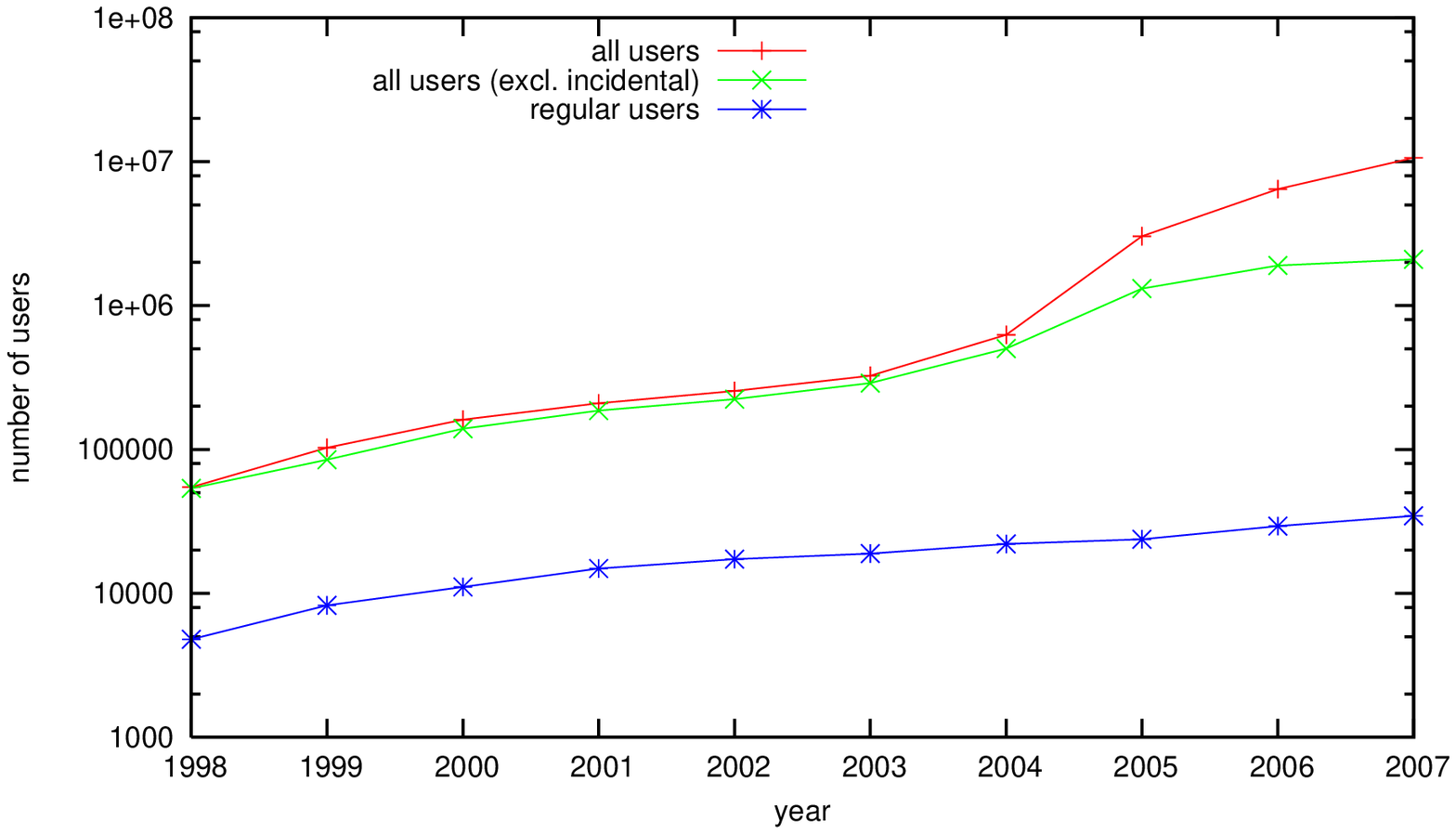}
  \plotone{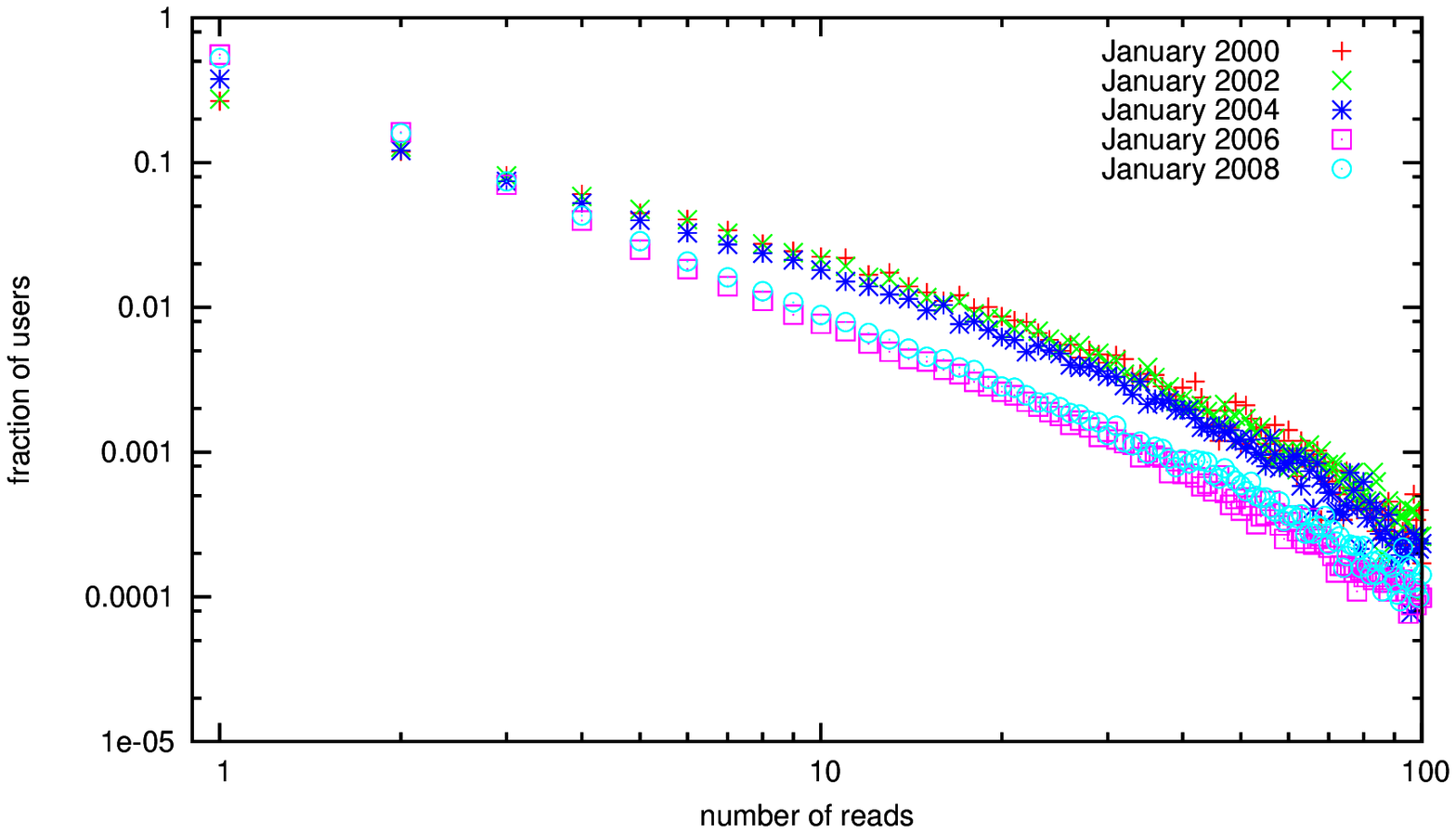}
  \caption{Top: Number of ADS users over time. The line marked '+' shows the total number of users. This includes incidental users who just look at an abstract. Excluding incidental users, the total number of users is shown by the line marked with '$\times$' (these users request additional metadata, besides abstracts, and perform queries). Finally, the number of users who use the ADS regularly (10 or more times per month) is given by the line marked with '*'. Bottom: user fraction ranked by the number of reads for the month of January in 2000, 2002, 2004, 2006 and 2008.}
  \label{fig:users}
\end{figure}

Figure~\ref{fig:users} (top) illustrates the observation we made in the introduction: "Since the ADS started to be indexed by general search engines, the number of incidental users
has increased dramatically". The line marked '+' shows the total number of users. This includes incidental users who just look at an abstract. Excluding incidental users, the total number of users is shown by the line marked with '$\times$' (these users request additional metadata, besides abstracts, and perform queries). Finally, the number of users who use the ADS regularly (10 or more times per month) is given by the line marked with '\GStar'. The bottom panel is an illustration of the qualitative change in use. It shows the distribution of the fraction of users as a function of the number of reads in the month of January in the years 2000, 2002, 2004, 2006 and 2008.  

Table~\ref{table:period08} gives an overview of the number of visitors entering through one of those external websites, for the period of January and February of 2008. The total number of visits in this period was 5,941,983.

\begin{table}
\begin{center}
\begin{tabular}{|l|l|}
\hline
\multicolumn{2}{|c|}{ADS Usage Through External Sites} \\
\hline
Site & Number of visits \\ \hline
Google & 1,920,797 \\
Google Scholar & 315,864 \\
Wikipedia & 31,094 \\
Astr. Pict of the Day & 23,742 \\
arXiv & 17,474 \\
MSN & 11,556 \\
Yahoo & 2,015 \\
Ask.com & 1,354 \\
\hline
\end{tabular}
\caption{Use of the ADS in January and Feburary, 2008.The total number of visits in this period was 5,941,983}
\label{table:period08}
\end{center}
\end{table}

\subsection{Readership for different types of readers}
\begin{figure}[!h]
  \plotone{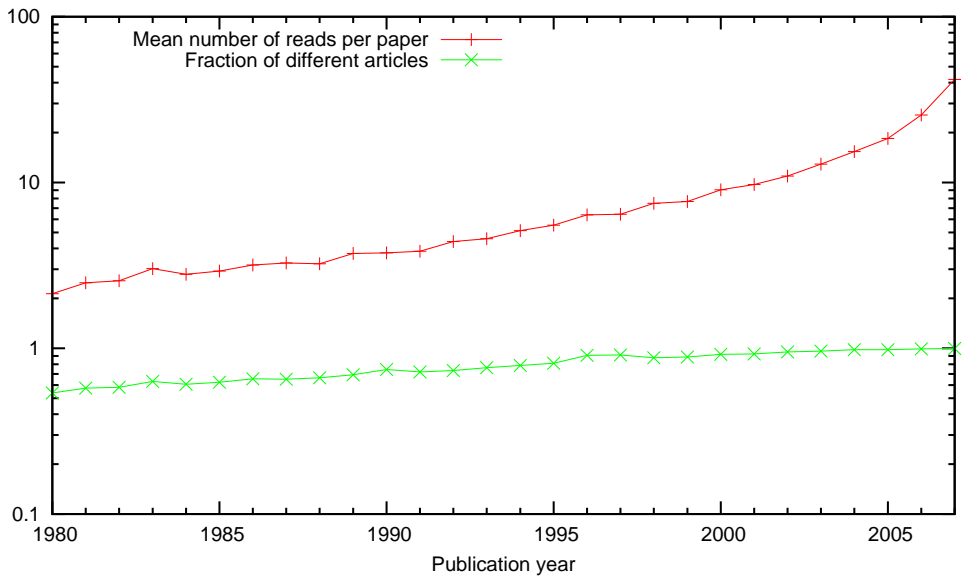}
  \caption{Article use through the ADS as a function of publication year}
  \label{fig:usage}
\end{figure}

\begin{figure}[!h]
  \plotone{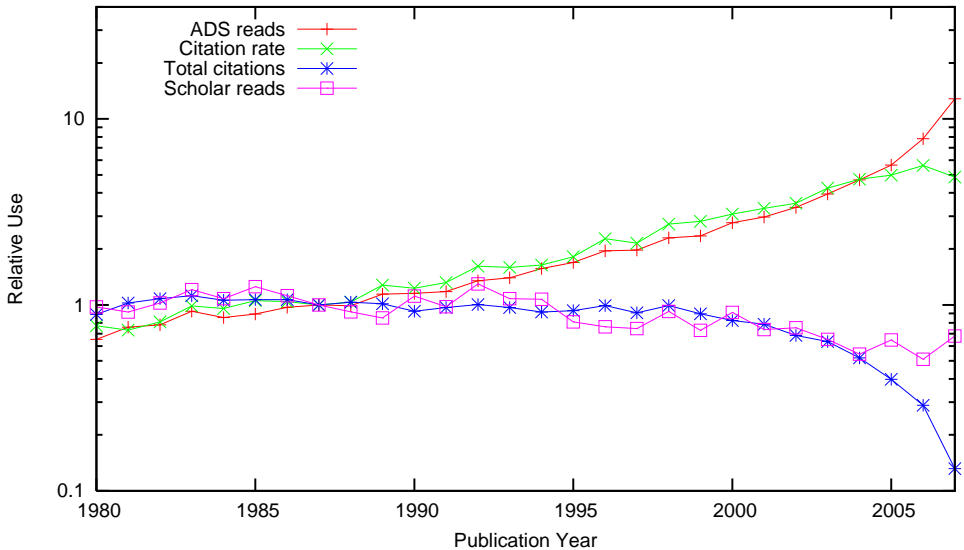}
  \caption{Comparison of readership patterns from ADS and Google Scholar queries, as observed in ADS's access logs. The line marked with '+' shows the readership use by people using the ADS search engine. The line marked with '$\Box$' corresponds with the readership use by people who used the Google Scholar engine. The line marked with '$\times$' shows the citation rate to the articles, while the line marked with '*' respresent their total number of citations.}
  \label{fig:types}
\end{figure}

For the first part of our analysis (figures~\ref{fig:usage} and~\ref{fig:types}) we have limited ourselves
to data requests for the following journals: {\it The Astrophysical Journal} (including
Letters, but excluding the Supplement), {\it The Astronomical Journal}, {\it The Monthly Notices of the Royal
Astronomical Society} and {\it Astronomy \& Astrophysics}. These journals constitute the core of research publications in astronomy, which all active astronomers read on a regular basis. First we determined the total use during these two months and the total number
of different articles for which data was requested. Since the number of
articles published is a function of time, it makes more sense to scale the
totals with the number of papers published in a given year. This gives the
mean usage for each publication year and the fraction of published papers for
which information was requested.

These numbers are shown in figure~\ref{fig:usage}. The line marked with '+' shows the mean usage per paper, and the line marked with '$\times$' represents the fraction of different articles for which data was requested. This figure is very similar to figure 12 in \citet{kurtz00}. 

What picture do we get when we zoom in on different types of users? In
particular, we will look at users who we will call ``ADS regulars'' (mostly
astronomers and physicists) and people requesting information through Google
Scholar. The group of ``ADS regulars'' consists of people who use the system
more than 10 times per month. Figure~\ref{fig:types} shows the mean usage for these
types of users, as function of publication year. The line marked with '+' shows the readership use by people using the ADS search engine. The line marked with '$\Box$' corresponds with the readership use by people who used the Google Scholar engine. The line marked with '$\times$' shows the citation rate to the articles, while the line marked with '\GStar' respresent their total number of citations. In other words, this figure compares the so-called "obsolescence functions" for cites and use (see e.g.~\citet{kurtz03}) for articles published in the four main astronomy journals as read by these two types of users.

An interesting metric for the ``ADS Regulars'' is the median of monthly usage. In the period of January 1998 through January 2008 the median for the monthly number of reads (by these users) turns out to be fairly constant at a value of 21$\pm$1 reads per month.

\subsection{Readership for different geographical regions}
\begin{figure}[!htp]
  \plotone{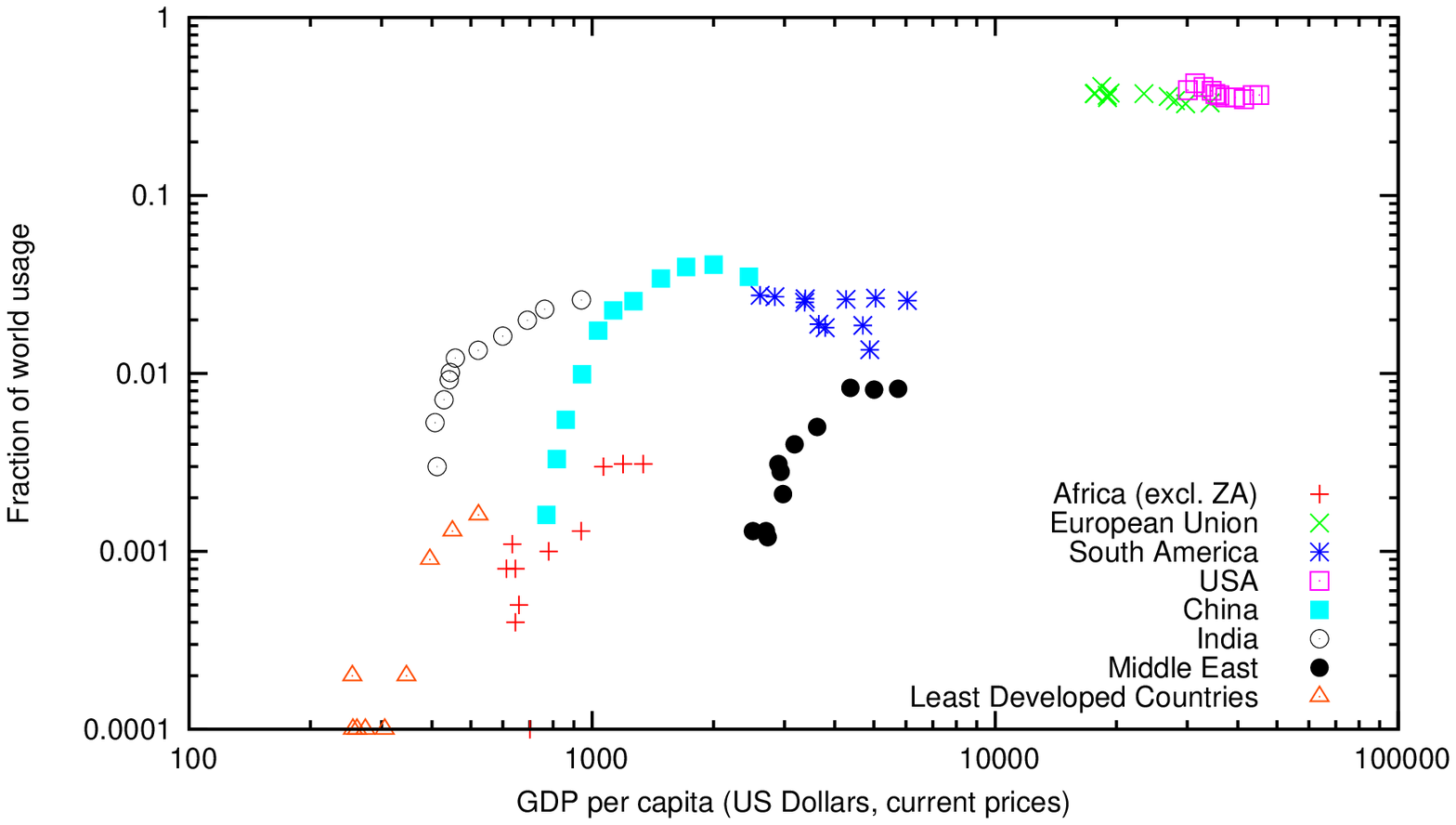}
  \plotone{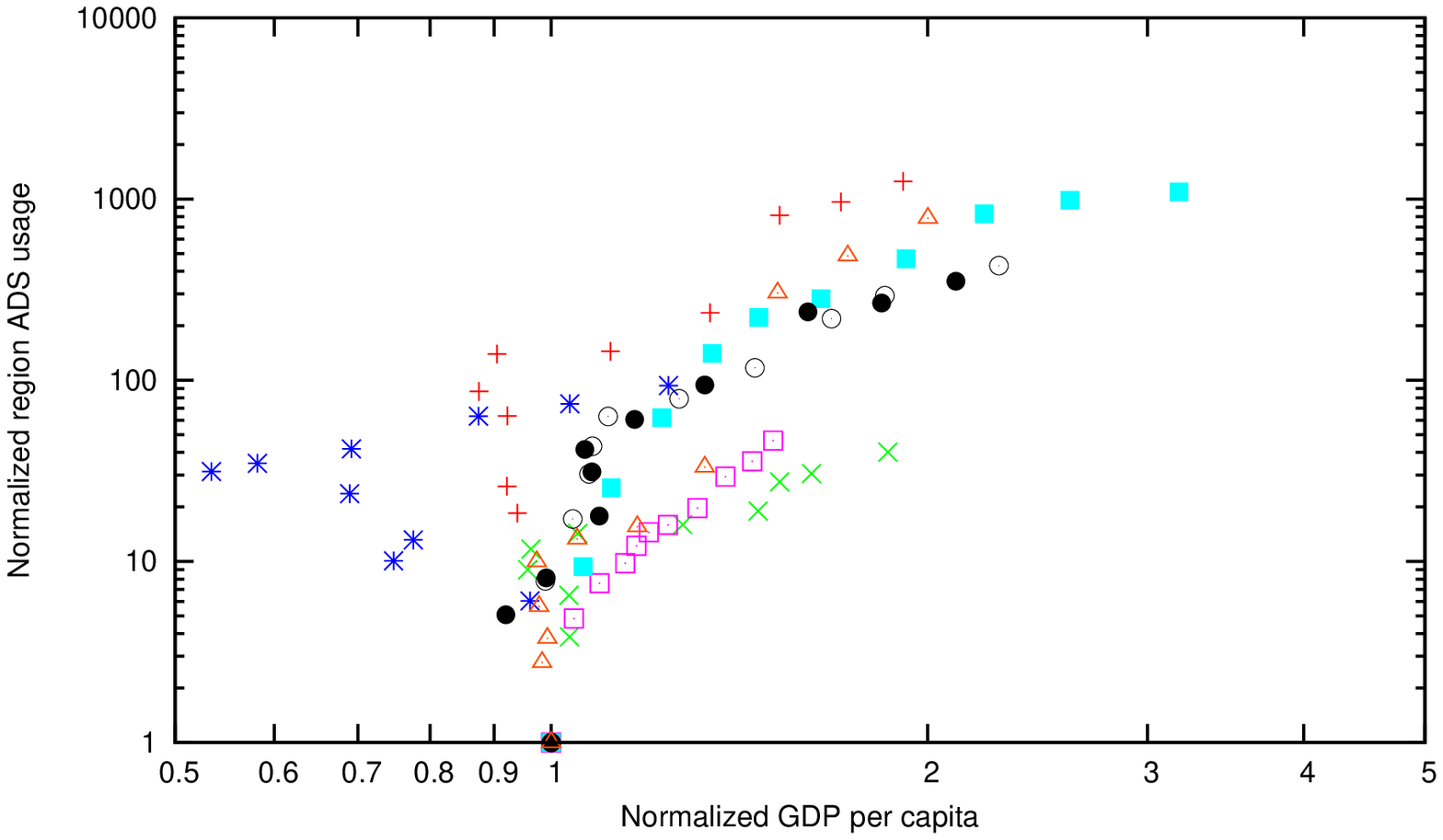}
  \plotone{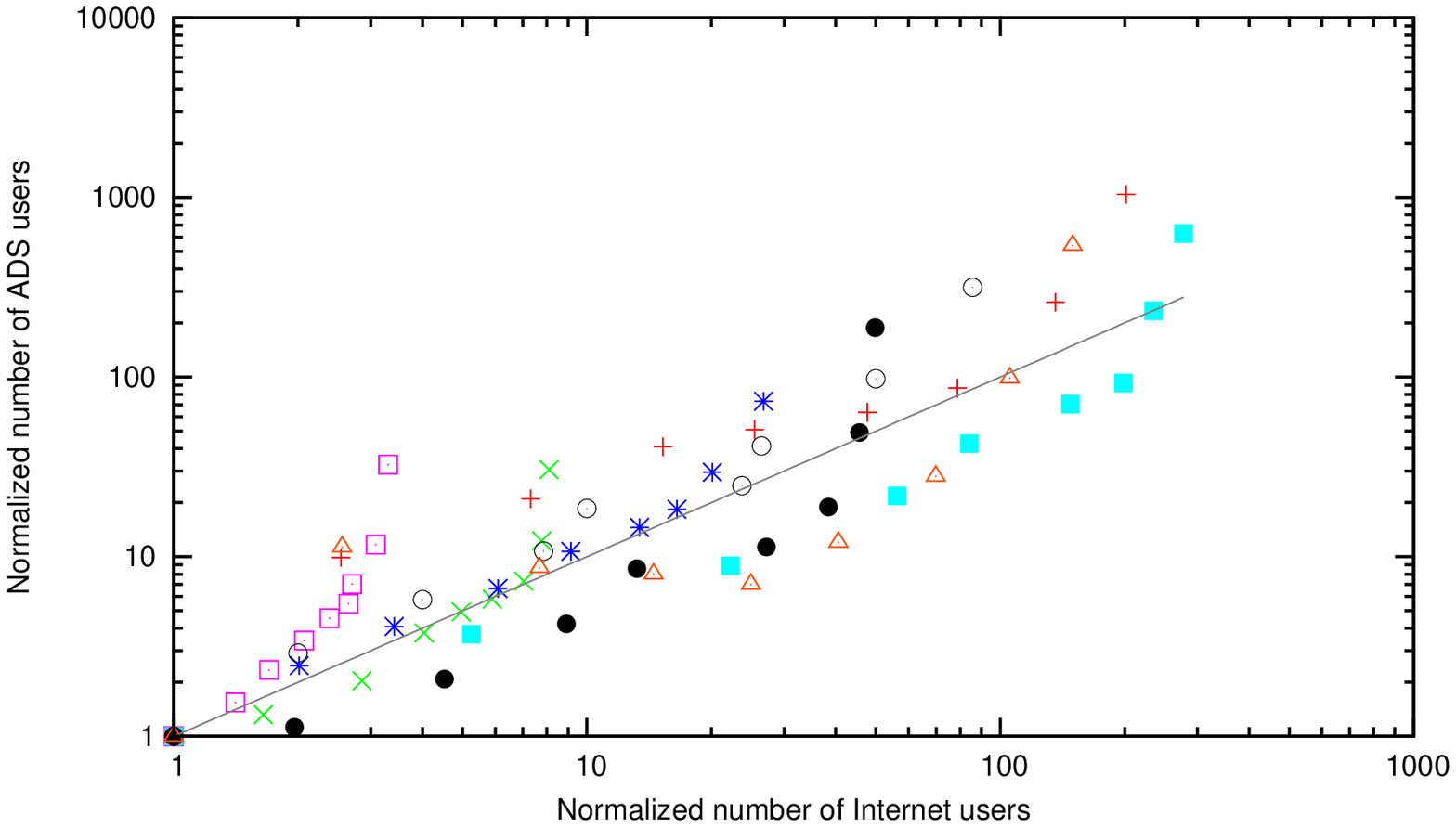}
  \caption{Top panel: Fraction of world usage as a function of GDP per capita (GPC) for 8 different regions in the time period of 1997 through 2007. Middle panel: Region ADS usage as a function of GPC. Quantities have been normalized by their value in the year 1997. Lower panel: Amount of ADS users in a region as a function of the number of Internet users for that region. Quantities have been normalized by their value in the year 1997. The solid line indicates the 1:1 line.}
  \label{fig:GDP}
\end{figure}

For the second part of our analysis we zoom in on usage data, to see how readership varies per geographical region. In the previous section, we mentioned that our data logs also record the origin of requests. This allows us to determine use as a function of geographical region. Since science and technology depend heavily on budgets, it is particularly interesting to look at the readership in a particular region as a function of GDP per capita (GPC), especially for developing countries. Figure~\ref{fig:GDP} shows the results for eight regions of various economic strength. Each data point corresponds with one year. 

In the top panel, we explore readership as a fraction of total readership in a given year. This fraction will tell us to what extent region usage growth follows the growth trend on world level. We will refer to the set of data points for each region as a ``trail''. A trail maps out the relation between GPC and the fraction of world usage over time. The trails shown in this panel can be classified into three groups. The trails for the EU and the USA evolve from the left to the right and slightly down, as time progresses. The trail for South America initially moves up (from the smallest fraction of world usage) and to the left, as the region moves into a recession, and then to the right, with a close to constant fraction of world usage. The trails for the remaining regions show an evolution of an increase in both GPC and the fraction of world usage. 

The middle panel looks at pure growth within the region as a function of GPC. In this figure usage has been normalized by the 1997 level, so the numbers show a relative growth with respect to 1997. Normalized, this plot will show similarities in intrinsic growth. The general flow in this diagram is up and to the right, as time progresses. The most pronounced exception is South America, moving into its deep recession. Essentially there seem to be two classes of trails, one formed by the EU and the USA and the other by the remaining regions. 

The lower panel compares the number of ADS users in a region with the number of Internet users, both normalized with their 1997 values. The flow of time here is to the right and up. Points above the solid line indicate that in a particular region the number of ADS users grows faster than the number of Internet users.

In order to get the data used in figure~\ref{fig:GDP} the following operations were performed on the origin information in our data logs: (a) requests originating from a ".com" or ".net" domain were assigned to the country of the referer URL, and (b) we set a limit of 2000 reads per year to any user (thus filtering out atypical usage). The region of ``Least Developed Countries'' consists of the 49 countries (as of the writing of this paper), as defined by the UN. The appendix includes the list of countries in this category.
\section{Discussion}

Figure~\ref{fig:users} shows how the ADS has become a public service whose reach goes well beyond the scholarly community. Since 2005, the number of people visiting the ADS via external sites has increased dramatically. This aspect will only intensify with the advent of the World Wide Telescope (\citet{gray02}), which has links to the ADS. Google Sky (\citet{scranton07}) already contains context-sensitive menus that enable positional searches for papers in ADS. Incidental users have contributed the most to the strong increase that started in 2005. We have observed that a fraction of these people actually started using the ADS on a semi-regular basis. The lower panel of figure~\ref{fig:users} illustrates the qualitative change in ADS usage. In January of 2006 and 2008, the distribution of reads over users (shown as fraction of the total number of users) is different from the other years shown. It shows that the ratio of frequent to infrequent users has changed considerably. Table~\ref{table:period08} shows that in the period of January and Febraury of 2008, almost 2 million visitors came to the ADS through Google, which seems to have become the trend. On average, these people get 2.2 abstracts of research grade papers each. We think that this has an impact on the science education of the general public, which may be compared to some general circulation magazines. Quantifying this would need further research.

Figure~\ref{fig:usage} reports on popularity of papers as a function of publication date. What papers (in terms of their age) get the most attention from people who use the ADS regularly? This figure shows that more recent papers get the most attention. This is not just because they are read more. The fact that the ADS reads increase more rapidly, for more recent papers, than the fraction of unique papers being read, shows that the fraction of interesting papers is larger in more recent papers. This was also observed in~\citet{kurtz00}.

Figure~\ref{fig:types} can be read in various ways. One interpretation is that use obsolescence can be substantially different for different types of users, even when accessing the same documents. How a search engine works, can have significant effects. Google Scholar use approximately matches the total number of citations, which is similar to the reading habits of students. Of course, ADS can also provide results similar to Google Scholar by requesting that articles ADS returns be sorted by citation count. There is an underlying reason for the strong correlation between the total number of citations and the readership patterns through Google Scholar. This reason is the correlation between the PageRank and the total number of citations (see e.g.~\citet{chen07}). If we classify papers on the basis of their references and citations and average the PageRank over these classes, it turns out that the average PageRank is proportional to the total number of citations and is independent of the number of references (see~\citet{fortunato06}). A consequence of the results shown in figure~\ref{fig:types} is that ADS provides what researchers want, while Google Scholar does not. 

Another interpretation of figure~\ref{fig:types} focuses on obsolescence of use and cites. A fundamental difference between cites and use is that the former is a public act, while the latter is a private act. Citations are created by authors of scholarly articles, while use, in general, is not solely the result of actions by authors. In other words, authors are often users, but there is a large set of users who are not authors. This contributes to the complexity of the relation between use and cites. The reader can find a detailed mathematical analysis of obsolescence in~\citet{egghe00} and a phenomenological analysis in~\citet{kurtz03}, where use is modeled as consisting of 4 modes, each with its own characteristic time scale. Just like the results in~\citet{kurtz03}, the results in this paper represent the mean current use per article published as a function of article age. In this way we directly measure the intrinsic decay. It is interesting to observe the close correlation between use and cites in the mean. It falls outside the scope of this paper to model our observations. We refer to~\citet{kurtz03} for a detailed discussion of obsolescence of cites and use. Actual citation distributions among papers classified according to popularity will need further research and will be the subject of future publications.

What does it mean to have a median that is fairly constant at a value of about 21 reads per month? It is an indication that all our frequent users on average use the ADS on a daily basis. Initially this meant that all professional astronomers use the ADS on a daily basis (see~\citet{kurtz05} and \citet{tenopir05}). This is probably still true, but with the addition of a growing group of physicists. The number of 21$\pm$1 reads per month is in agreement with the findings in~\citet{kurtz05} (where a median of 22 reads per month was found for the month of August in 2001).

Figure~\ref{fig:GDP} tells many stories. In the top and middle diagrams, each region has 11 datapoints, corresponding to one year (1997 through 2007). The lower diagram has 9 points for each region (1997 through 2005). South Africa (ZA) was omitted from the region "Africa", because it would completely dominate the analysis for this region (the fraction of world usage for South Africa alone is more than for the rest of Africa).

In the top panel of figure~\ref{fig:GDP} it is immediately clear that the EU and the USA are different from the other regions. Obviously, for both, the GPC will be higher than the other regions. Since the lion's share of astronomy research is performed in either the USA or the EU, it is also no surprise that their fraction of world usage is substantially higher than for the other regions. Significant is the difference in how the fraction of world usage changes over time. For both the EU and the USA, this fraction gradually decreases over time. The explanation for this is that these two regions represent the bulk of ADS usage (ranging from 84\% of world usage in 1997 to 70\% in 2007). ADS usage comes from existing Internet users, because in these regions, the number of Internet users is more or less saturated. So, even though there is a steady increase in the number of users in these regions, the fraction decreases because the overall usage increases faster. China displays a similar trend in the recent past, following a period of fast growth (until 2004). The character of the Chinese economy has changed: initially we see growth typical for a low-income region, but probably around 2004, China moved into being a middle-income region. This seems to be the case if we take the number of Internet users as an indicator for economic growth. The number of Internet users in middle- and high-income regions saturates over time (see below). The data points for South America show that there is an increase in ADS usage, even when the economy for that region goes through heavy recession. Around 2002 the fraction settles on a value of about 3\%. 

\begin{figure}[!h]
  \plotone{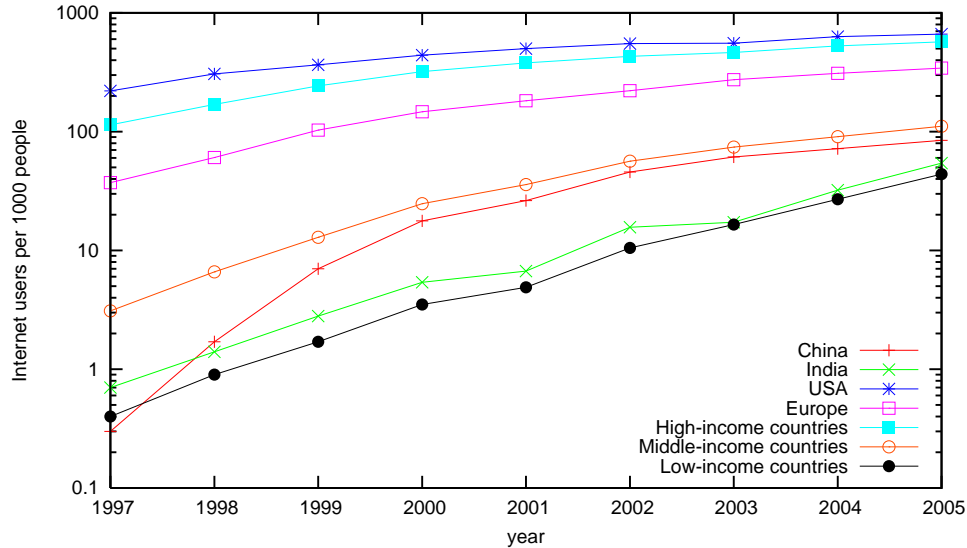}
  \caption{Internet users per 1000 people for China, India, USA, Europe and high-, middle and low-income countries (source: WRI, EarthTrends database)}
  \label{fig:Internet}
\end{figure}

The middle panel of figure~\ref{fig:GDP} shows that all regions display an intrinsic growth (with respect to their 1997 values). The overall trend is that the intrinsic growth in the EU and USA is slower than in the other regions, which is to be expected for regions where the density of Internet users changes relatively little (because of the wide initial penetration of Internet connectivity). Figure~\ref{fig:Internet} illustrates this fact.
The definitions of high-, middle- and low-income countries are those of the Worldbank: high-income countries are those countries with a Gross National Income per capita (GNC) in 2007 of \$11,456 or more, while middle-income countries are those countries with a GNC between \$936 and \$11,455 and low-income countries are countries with a GNC of less than \$935. This figure shows that, using Internet usage as indicator, China changed from being a low-income country to a middle-income country. Low-income countries do not show a clear flattening trend in the Internet users density, which is clearly present for middle- and high-income countries. This is probably due to the fact that low-income countries still have a lot of potential for growth. China becoming a middle-income country is probably the reason for the decrease in the fraction of world-usage, seen in figure~\ref{fig:GDP} (top).

After the Internet user density flattens, the ADS user density still increases, because there is still substantial potential for use diversification within the existing body of users. This is clearly shown in the lower panel of figure~\ref{fig:GDP}, most prominently for the USA. The number of ADS users increases rapidly, while the number of Internet users does not. Once everybody is online, the only thing that changes is browsing behavior. The EU shows a similar trend, with a delay. This is probably due to a longer penetration time of the Internet in various member states of the EU. In general, points above the solid line in the lower panel of figure~\ref{fig:GDP} indicate a growth of the number of ADS users that is faster than the growth of Internet users in that region.

\section{Conclusions}

In terms of its audience, the ADS has not only changed quantitatively, but also qualitatively. Besides a steady growth of the ADS regular users, we observe a dramatic increase in incidental users. The ADS is per definition the gateway to online literature for scientists, used by virtually all professional astronomers on a daily basis. Since 2005 there is a growing role as a source of science education of the general public.

Comparing the group of ``ADS regulars'' with the group visiting the ADS via Google Scholar shows that the obsolescence curve for the latter is fairly flat, corresponding with reading behavior by people acquainting themselves with a subject. This means Google Scholar is not the right tool for staying up-to-date with the latest events in a field. Looking at how professional astronomers use the ADS shows that the obsolescence function for them closely follows the citation rate (as a function of paper age).

Although ADS usage increased in regions like the EU and the USA, the percentage of world usage has decreased for these regions. This is because the growth in World usage is mainly driven by regions with the biggest potential for growth. The density of Internet users reaches a saturation point in middle- and high-income regions at which point ADS usage increases at a slower rate. It is encouraging to see the rapid increase in Internet user density in low-income regions and a similar increase in the number of ADS users in those regions. It indicates that increased access to electronic information is being used and in this sense there is a narrowing of the ``Digital Divide'' for these regions. Whether this increased access also resulted in an increased scientific output needs further bibliometric research.


\section{Appendix}

A country is classified as a ``Least Developed Country'' when it meets the following three criteria (\citet{UN07}, \citet{UN08}):
\begin{itemize}
\item a low-income criterion, based on a three-year average estimate of the gross national income (GNI) per capita (under \$745 to be included in the list, above \$900 to be removed from the list);
\item a human capital status criterion, involving a composite Human Assets Index (HAI) based on indicators of: (a) nutrition: percentage of population undernourished; (b) health: mortality rate for children aged five years or under; (c) education: the gross secondary school enrolment ratio; and (d) adult literacy rate; and
\item an economic vulnerability criterion, involving a composite Economic Vulnerability Index (EVI) based on indicators of: (a) population size; (b) remoteness; (c) merchandise export concentration; (d) share of agriculture, forestry and fisheries in gross domestic product; (e) homelessness owing to natural disasters; (f) instability of agricultural production; and (g) instability of exports of goods and services.
\end{itemize}
To be added to the list, a country must satisfy all three criteria. In addition, since the fundamental meaning of the LDC category, i.e. the recognition of structural handicaps, excludes large economies, the population must not exceed 75 million. To become eligible for graduation, a country must reach threshold levels for graduation for at least two of the aforementioned three criteria, or its GNI per capita must exceed at least twice the threshold level, and the likelihood that the level of GNI per capita is sustainable must be deemed high.

\begin{table}
\begin{tabular}{ l l l }
\hline
\multicolumn{3}{l}{\bf{Africa (33)}} \\
Angola & Ethiopia & Niger \\
Benin & Gambia & Rwanda  \\
Burkina Faso & Guinea &  S\~ao Tom\'e and Pr\'incipe \\
Burundi & Guinea-Bissau &  Senegal \\
Central African Republic & Lesotho & Sierra Leone \\
Chad & Liberia &  Somalia \\
Comoros & Madagascar &  Sudan \\
Dem. Rep. of the Congo & Malawi & Togo \\
Djibouti & Mali &  Uganda \\
Equatorial Guinea & Mauritania & United Rep. of Tanzania \\
Eritrea & Mozambique & Zambia \\
\multicolumn{3}{l}{\bf{Asia (15)}} \\
Afghanistan & Lao People's Dem. Rep. & Solomon Islands \\
Bangladesh & Maldives & Timor-Leste \\
Bhutan & Myanmar & Tuvalu \\
Cambodia & Nepal & Vanuatu \\
Kiribati & Samoa & Yemen \\
\multicolumn{3}{l}{\bf{Latin America and the Caribbean (1)}} \\
Haiti & & \\
\hline
\end{tabular}
\caption{Least Developed Countries (UN definition)}
\label{table1}
\end{table}

\end{document}